\documentclass{aa}  
\usepackage[varg]{txfonts}
\usepackage{natbib}
\bibpunct{(}{)}{;}{a}{}{,}
\usepackage{tabularx}
\usepackage{booktabs}
\usepackage{inputenc}
\usepackage{textcomp}
\usepackage{subfigure}
\usepackage{amsmath}
\defcitealias{rhee_2017}{R17}
\defcitealias{pasquali_2019}{P19}

\begin{document}

\title{Estimating infall times of galaxies around clusters: how accurately can it be done with observational data?}

% \author{
% Haoran Dou (窦浩然)\inst{1} \and 
% Heng Yu (余恒)\inst{1}\thanks{Corresponding author: \email{yuheng@bnu.edu.cn}}
% }
\author{
Haoran Dou\inst{1} \and 
Heng Yu\inst{1}\thanks{Corresponding author: \email{yuheng@bnu.edu.cn}}
}

\institute{School of Physics and Astronomy, Beijing Normal University Beijing 100875, PR China}

\titlerunning{Estimating galaxy infall time in the $R-V$ diagram}
\authorrunning{Haoran Dou \and Heng Yu}

\abstract
{The environment plays a crucial role in galaxy evolution, particularly for galaxies infalling into clusters.
Accurately estimating the infall times of galaxies from observations can significantly enhance our understanding of the environmental effects on galaxy evolution. }
{This paper aims to evaluate existing methods for estimating infall times via the $R-V$ diagram, explore possible strategies to improve accuracy in estimating infall times, and discuss fundamental limitations.}
{We utilize a TNG300-1 simulation and construct the $R-V$ diagram that is directly comparable to the observations.
Using the same dataset, we systematically compare four commonly used methods, including the projected radii, caustic profiles, and two discrete methods.
A simple linear partition is also considered as a reference. }
{Each method exhibits distinct characteristics.
While the linear partition slightly outperforms other methods, all methods suffer from limited accuracy ($\gtrsim 2.6$ Gyr), constrained by the intrinsic dispersion ($2.53$ Gyr) of infall times in the $R-V$ diagram.
Given this limit, we explore two potential approaches that can improve accuracy: (1) the infall time dispersion is smaller in more dynamically relaxed clusters, and (2) employing two estimates of infall times instead of one reduces the dispersion to $\lesssim1.5$ Gyr.
We further demonstrate that the intrinsic dispersion primarily arises from orbital overlap: galaxies in different orbital phases overlap with each other in the $R-V$ diagram and thus appear indistinguishable. }
{Orbital overlap fundamentally limits the accuracy of infall time estimation.
The linear partition approach could be a simple and robust estimation.}

\maketitle

\section{Introduction} \label{sec:intro}

The evolution of galaxies is one of the fundamental questions in astrophysics.
It is widely known that the environment plays an important role in galactic evolution \citep[e.g.,][]{Dressler_1980,Dressler_1997,Goto_2003,Kauffmann_2004,Alpaslan_2015,Chen_2017,Taylor_2023,Shi_2024,Zheng_2024}.
Galaxies residing in denser environments, such as clusters and groups, tend to be red, gas-depleted, metal-rich, of early-type morphology, and with lower star formation rates (SFRs), in contrast to field galaxies, which typically exhibit opposite properties.

Tracing the infall process of galaxies into clusters offers a promising way to understand how the environment drives galaxy evolution.
In simulations, full six-dimensional kinematic information and complete motion histories allow for accurate reconstruction of the orbits of galaxies \citep[e.g.,][]{Mahajan_2011,Oman_2013,Oman_2016,rhee_2017,pasquali_2019,Rhee_2020}.
However, observational data provide only three dimensions of kinematic information: two projected positions on the sky and one line-of-sight (LOS) velocity.
Therefore, the galaxy infall process is commonly analysed via the $R-V$ diagram, also referred to as the projected phase space (PPS) in simulation studies. This diagram comprises the projected cluster-centric radius ($R_{\rm 2D}$) and the LOS velocity relative to the cluster centre ($V_{\rm los}$).

Several studies have explored the potential of the $R-V$ diagram for tracing the infall process and have proposed various methods for estimating galaxy infall times.
\citet{Noble_2013,Noble_2016} employed caustic profiles to delineate regions corresponding to different infall stages.
\citet{Oman_2013} and \citet{Oman_2016} used the Multidark Run 1 dark matter-only simulation to construct orbital libraries and calculated probability density functions of infall times at each location in the diagram.
\citet{rhee_2017} divided the $R-V$ diagram into five zones designed to maximize the fraction of galaxies belonging to a specific infall population in each zone.
\citet{pasquali_2019} introduced a set of quadratic curves to partition the $R-V$ diagram into eight zones, achieving improved alignment with the mean infall time distribution compared with caustic profiles.
These methods have been widely adopted in observational studies \citep[e.g.,][]{Kim_2023,Sampaio_2021,Brambila_2023,Oxland_2024,Sampaio_2024}.

However, these methods have not been systematically evaluated under the same conditions. It remains unclear which method provides the most reasonable estimates of infall times or to what extent infall times can be accurately estimated from observational $R-V$ diagrams.
Although some works have reported accuracy levels, such as 2.58 Gyr in \citet{Oman_2013} and $1.49\sim2.60$ Gyr in \citet{pasquali_2019}, there is no easy way to combine the intrinsic uncertainties with the observational data.
This study aims to address that question through a consistent and comparative evaluation of different methods while also exploring possible improvements.

The simulation data used for this study are presented in Section \ref{sec:data}, followed by a detailed description of the infall time distribution in the $R-V$ diagram in Section \ref{sec:infall_time}.
Section \ref{sec:comparison} compares four commonly used methods for infall time estimation alongside a simple linear partition method for reference.
In Section \ref{sec:discussion}, we explore possible approaches to improve accuracy, and in particular, we discuss the orbital overlap issue that contributes to the main dispersion in Section \ref{sec:orbit_overlap}.
Finally, we summarize the results in Section \ref{sec:summary}.

\section{Data} \label{sec:data}

\begin{figure*}[htb]
    \centering
    \includegraphics[width=0.95\linewidth]{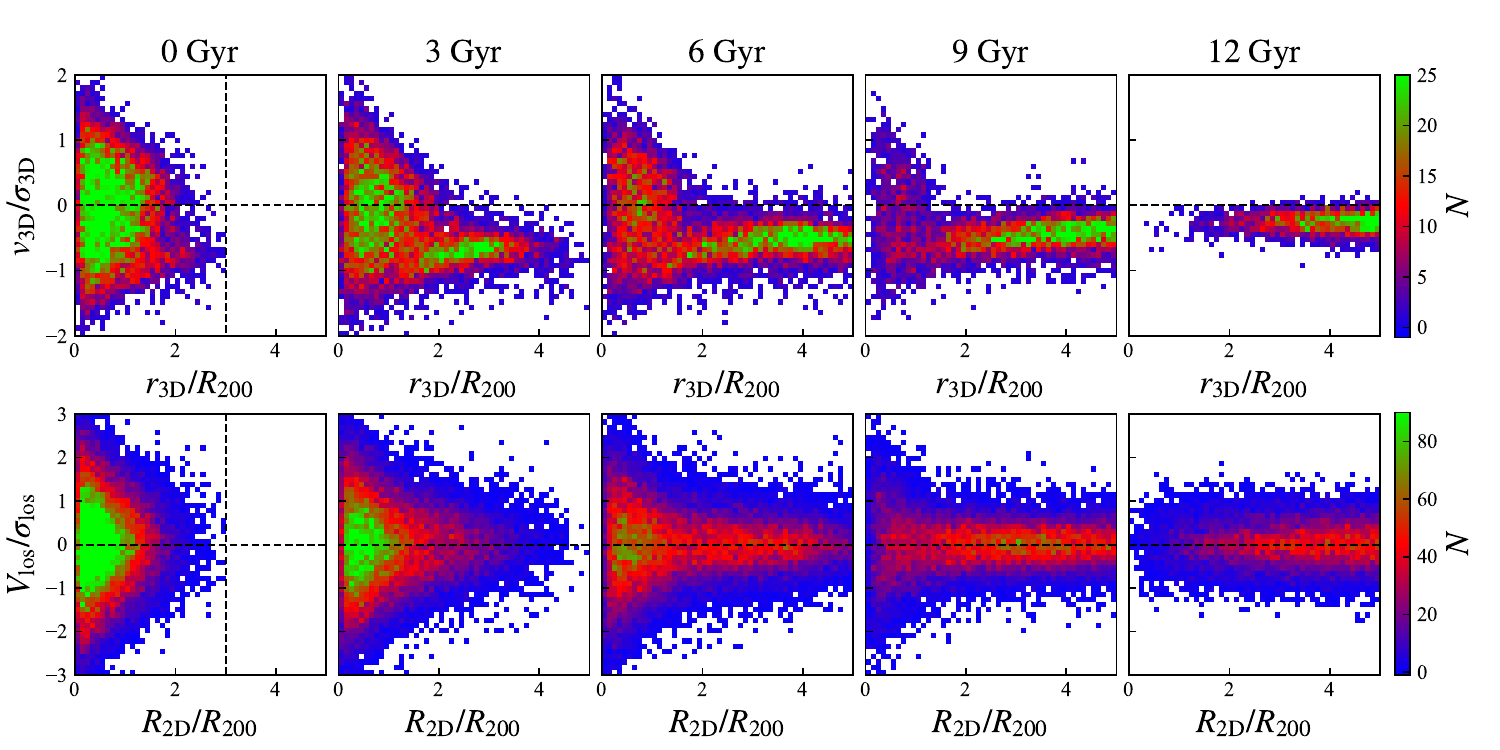}
\caption{Number distributions of the galaxies in the phase space and the $R-V$ diagram at five different lookback times.
The vertical lines in the leftmost panels represent the $3R_{200}$ boundary for member galaxies.}
    \label{fig:infall_his}
\end{figure*}

\subsection{Simulations}

IllustrisTNG (hereafter, TNG) is a suite of large-volume, cosmological, gravo-magnetohydrodynamic simulations that includes a comprehensive model for galaxy formation physics \citep{Springel_2018,Pillepich_2018,Naiman_2018,Nelson_2018,Marinacci_2018}.
TNG 300 has the largest volume (${205\ h^{-1} {\rm Mpc}}^3$) of TNG, and TNG 300-1 has the highest resolution level of TNG 300, including initial $2\times2500^3$ particles, with a dark matter cell mass of $4.0\times10^7h^{-1}{\rm M}_\odot$, and a baryonic matter cell mass of $7.6\times10^6h^{-1}{\rm M}_\odot$.

The public release of TNG 300 provides 100 snapshots between $z=20$ and $z=0$,
and a group catalogue is provided for each snapshot.
These catalogues contain halos (groups or clusters) identified with the standard Friends-of-Friends algorithm \citep{Davis_1985}, as well as subhalos (galaxies) identified with the Subfind algorithm \citep{Springel_2001,Dolag_2009}.
The merger histories of subhalos are constructed via SubLink \citep{Rodriguez_2015} and LHaloTree \citep{Springel_2005}.
Hereafter, we use "clusters" and "galaxies" to refer to halos and subhalos, respectively.

\subsection{Cluster sample}

We select clusters from the redshift $z=0$ snapshot, applying a lower mass threshold of $M_{200} > 10^{14}{\rm M}_\odot$.
As \citet{rhee_2017} demonstrated that galaxy behaviour in the $R-V$ diagram shows negligible dependence on cluster mass, we exclude systems with $M_{200} < 10^{14}{\rm M}_\odot$, which are typically classified as 'groups'.
To minimize the impact of cluster mergers, which can complicate member galaxy motions and introduce additional dispersion into the infall time estimation, we require the clusters to have all member galaxies within $3R_{200}$.

While TNG provides merger trees and motion histories for galaxies, it does not explicitly track clusters.
Therefore, we use the central galaxies as proxies for the positions and motions of clusters.
In most cases, a central galaxy always remains at the centre of a cluster.
However, temporary misclassifications may occur in some snapshots, especially at high redshifts.
Although such misclassifications generally do not affect the overall properties of clusters, they can introduce discontinuities in the infall trajectories of member galaxies.
A cluster is excluded from our sample if its central galaxy is misclassified in more than three consecutive snapshots, as this would cause a significant disruption in the reconstructed trajectories.
If the misclassified snapshots are non-consecutive or no greater than 3, we simply discard those specific snapshots to retain as many clusters as possible.

Our final sample comprises 136 clusters, with $M_{200}$ ranging from $1.01 \times 10^{14}{\rm M}_\odot$ to $1.03 \times 10^{15}{\rm M}_\odot$.
Their characteristic radius ($R_{200}$) and mass ($M_{200}$) are obtained from the TNG group catalogues. 
$R_{200}$ is the radius within which the average matter density is 200 times the critical density of the universe at the corresponding redshift, and $M_{200}$ is the mass enclosed within $R_{200}$.

\subsection{Galaxy sample}

We select member satellite galaxies associated with the 136 clusters from the redshift $z = 0$ snapshot, excluding central galaxies from our analysis.
Galaxies that have undergone a temporary infall, meaning that they passed through a cluster once and subsequently exited, are also excluded.
To ensure that the galaxies are well resolved, only those with stellar masses $M_*>10^{9}\ {\rm M}_\odot$ are preserved.
After applying these criteria, the sample comprises a total of 10,087 galaxies.

\subsection{$R-V$ diagram}

To construct the observable $R-V$ diagram from simulation data, we align the LOS direction with the \textit{x}, \textit{y}, and \textit{z} axes in turn, thereby tripling the sample size to 408 clusters and 30,261 galaxies.
The LOS velocity is the component of velocity along the line of sight.
The projected radius is defined as the distance from a galaxy to the central galaxy on the projection plane.
For example, when taking \textit{x} axis as LOS, the LOS velocity is $v_x$, and the projected radius is:
\begin{equation}
    R_{\rm 2D}=\sqrt{(y-y_0)^2 + (z-z_0)^2}
\end{equation}
where the subscript 0 denotes the coordinates of the central galaxy.
To stack all clusters together, the projected radii are normalized by $R_{200}$ of host clusters, and LOS velocities are normalized and the LOS velocity dispersions ($\sigma_{\rm los}$), which are the standard deviations of the LOS velocities of member galaxies.
In the mean while, the Hubble flow correction is implemented as follow:
\begin{equation}
    V_{\rm los}=\left| v_i+H_0\times r_i \right|
\end{equation}
where $i$ represents the specific LOS direction (\textit{x}, \textit{y}, or \textit{z}), while $v_i$ and $r_i$ are the corresponding components of clustercentric velocity and radial vectors, respectively.

Moreover, the 3D phase space (hereafter referred to as `phase space') serves as a reference for understanding the behaviour of galaxies.
It is defined by the normalized 3D radius ($r_{\rm 3D}/R_{200}$) and the radial velocity ($v_{\rm 3D}/\sigma_{\rm 3D}$).

For brevity, in the rest of this paper, we use lowercase $r$ and $v$ to denote $r_{\rm 3D}/R_{200}$ and $v_{\rm 3D}/\sigma_{\rm 3D}$, and uppercase $R$ and $V$ to denote $R_{\rm 2D}/R_{200}$ and $V_{\rm los}/\sigma_{\rm los}$, respectively.

\section{Infall time distribution in the $R-V$ diagram} \label{sec:infall_time}

\subsection{Infall process}

By tracing the historical positions and velocities of the member galaxies along their main merger trees, we illustrate their overall infall process in Fig.\ref{fig:infall_his}, where the galaxy number distributions are shown in both the phase space and the $R-V$ diagram at five lookback times.

A clear infall trajectory is visible in the phase space (upper panels of Fig.\ref{fig:infall_his}).
Galaxies initially fall from the outskirts (right) to the centre (left), with radial velocities directed inwards (negative).
Upon reaching the pericentre, the radial velocities become positive as the galaxies move outwards.
Then, the galaxies splash back to the outer region, and their velocities approach zero near the apocentre.
This cycle repeats, with orbital radii gradually shrinking until the galaxies eventually become virialized in the core region.
In the $R-V$ diagram (lower panels), a similar pattern is observed, although the distinction between the infalling and virialized regions is less pronounced due to projection effects.
This infall process is consistent with previous studies \cite[e.g.,][]{Mahajan_2011, Oman_2013, rhee_2017}.

Moreover, the galaxy distribution in the $R-V$ diagram is symmetric about the x-axis.
Assuming that galaxies are isotropically distributed around the cluster, each velocity component should follow a Gaussian profile centred on the average velocity.
To simplify the analysis, we use only the absolute value $|V|$ in the $R-V$ diagram in the remainder of this paper.

\subsection{The infall time}

\begin{figure}[ht]
    \centering
    \includegraphics[width=0.99\linewidth]{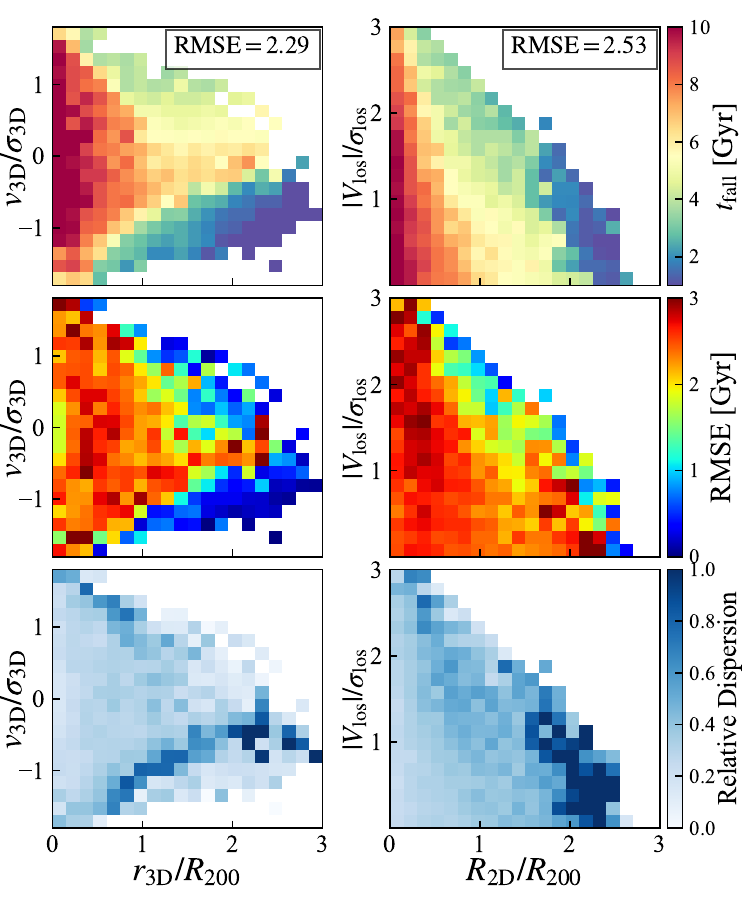}
\caption{Top: Median $t_{\rm fall}$ distributions in the phase space (left) and the $R-V$ diagram (right).
Middle: Dispersion distributions, defined as $\mathrm{RMSE}$ relative to the median in each pixel.
Bottom: Relative dispersion distribution, defined as the ratio of $\mathrm{RMSE_{pix}}$ to the median $t_{\rm fall}$. }
    \label{fig:tinf}
\end{figure}

We define the infall time ($t_{\rm fall}$) of a galaxy as the lookback time when it first crosses $3R_{200}$ of the cluster.
This boundary is sufficiently large to assign an infall time to each member galaxy and to ensure that no galaxy orbit exits.
The specific choice of boundary primarily affects the absolute value of $t_{\rm fall}$ (for example, when adopting $R_{200}$ as the boundary, $t_{\rm fall}$ values will be $\sim$2 Gyr smaller) but has little impact on the relative relationship between $t_{\rm fall}$ and the position in the $R-V$ diagram.
Therefore, slight variations in the definition of $t_{\rm fall}$ do not alter our overall results and conclusions.

The median $t_{\rm fall}$ distributions in both the phase space and the $R-V$ diagram are shown in the upper panels of Fig.\ref{fig:tinf}.
It is clear that galaxies with velocities and radii closer to zero tend to have larger $t_{\rm fall}$ values.
In the phase space, galaxies with $t_{\rm fall}<2$ Gyr are concentrated in a region characterized by large radii and negative radial velocities.
This finding is consistent with the infall trajectories shown in Fig.\ref{fig:infall_his}.
We refer to these galaxies as `recent-infall galaxies'.
In the $R-V$ diagram, recent-infall galaxies are also concentrated in the outer region with large radii.

To evaluate the accuracy of using the median values in pixels to estimate true infall times, we calculate the overall root mean square error ($\mathrm{RMSE}$):
\begin{equation}
    \mathrm{RMSE}=\sqrt{ \frac{1}{N_\mathrm{tot}}\sum_{i=1}^{N_\mathrm{tot}}(t_i - t_{\rm median,pix})^2  }
\end{equation}
where $N_\mathrm{tot}$ is the total number of galaxies, $t_i$ represents the true infall times of individual galaxies, and $t_{\rm median,pix}$ is the median infall time of galaxies in the corresponding pixel.
The overall $\mathrm{RMSE}$s reflect the fundamental dispersion in estimating infall times via kinematic data.
As shown in the upper panels of Fig.\ref{fig:tinf}, the dispersion in the $R-V$ diagram (2.53 Gyr) is greater than that in the phase space (2.29 Gyr), primarily because of the projection effect.

Then, we analyse the detailed dispersion distribution by calculating $\mathrm{RMSE_{pix}}$, which is similar to $\mathrm{RMSE}$ but averages the error of galaxies in individual pixels:
\begin{equation}
    \mathrm{RMSE_{pix}}=\sqrt{ \frac{1}{N_\mathrm{pix}}\sum_{i=1}^{N_\mathrm{pix}}(t_i - t_{\rm median,pix})^2  }
\end{equation}
where $N_\mathrm{pix}$ is the number of galaxies in the corresponding pixel.
The resulting $\mathrm{RMSE_{pix}}$ distributions are displayed in the middle panels of Fig.\ref{fig:tinf}.
Consistent with the overall $\mathrm{RMSE}$s, the dispersion is greater than 2 Gyr in most regions and even exceeds 3 Gyr in the inner region of the $R-V$ diagram.
Only the region of recent-infall galaxies in the phase space exhibits low dispersion of $\mathrm{RMSE}<1$ Gyr.

We also examine the relative dispersion, defined as the ratio of $\mathrm{RMSE_{pix}}$ to the median $t_{\rm fall}$, as shown in the bottom panels of Fig.\ref{fig:tinf}.
In the phase space, recent-infall galaxies exhibit low relative dispersions, as they have low absolute dispersions.
The largest relative dispersions exist in a transition region between recent-infall galaxies and inner galaxies, corresponding to the region with a median $t_{\rm fall}$ of approximately $2\sim4$ Gyr.
In the $R-V$ diagram, recent-infall galaxies suffer from the largest relative dispersions because of their high absolute dispersions and small infall times.

\section{Evaluating methods for estimating infall times} \label{sec:comparison}

Here, we compare several methods for estimating infall times in the $R-V$ diagram. They can be divided into two categories: continuous tracers and discrete zones.

\subsection{Continuous methods} \label{sec:conti}

\begin{figure*}[htb]
    \centering
    \includegraphics[width=0.95\linewidth]{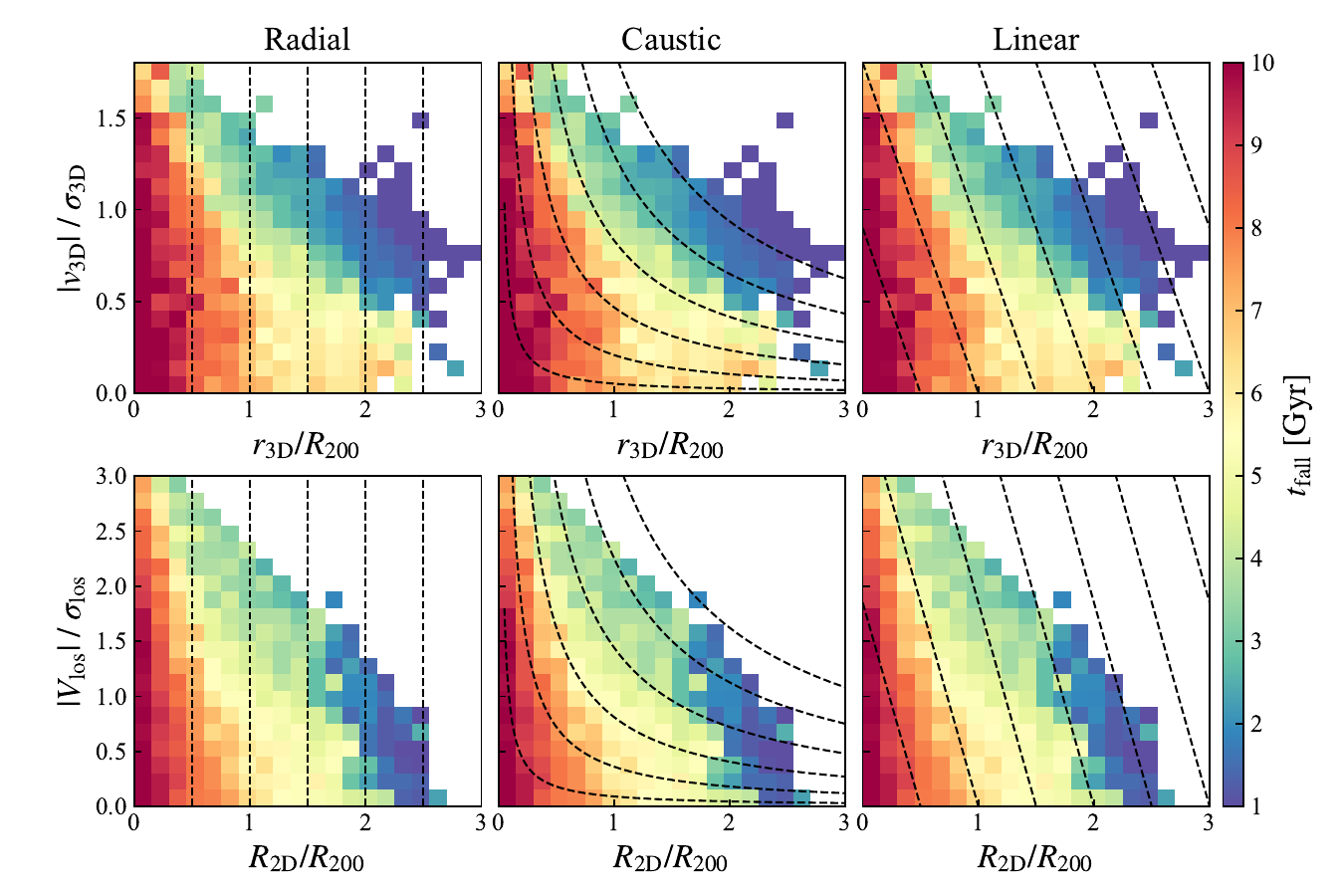}
\caption{Median $t_{\rm fall}$ distributions in the folded phase space (top) and the $R-V$ diagram (bottom). The three columns and corresponding dashed curves represent the projected radii, caustic profiles, and linear partitions, respectively.}
    \label{fig:continuous_map}
\end{figure*}
\begin{figure}[htb]
    \centering
    \includegraphics[width=0.8\linewidth]{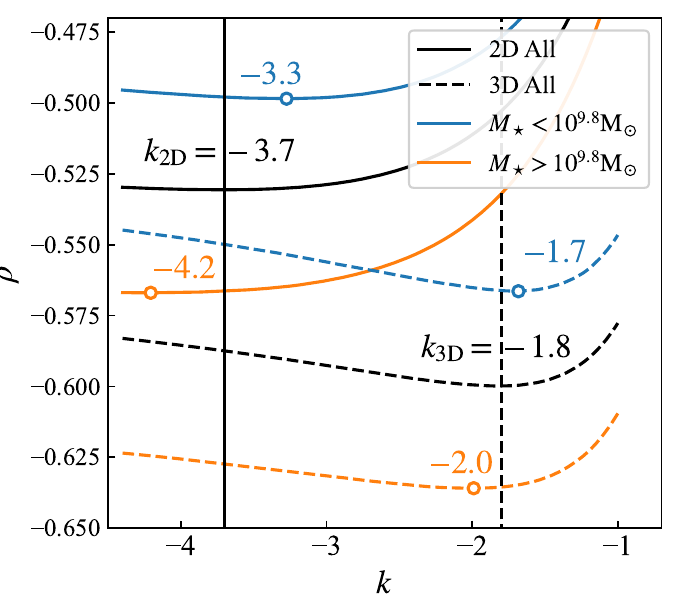}
    \caption{Spearman coefficients $\rho$ between infall time $t_{\rm fall}$ and the linear distance $d_{\rm linear}$ as functions of the slope $k$. 
    Solid and dashed curves show results in the $R$–$V$ diagram and phase space, respectively.
    Black curves represent the full galaxy sample, with vertical lines marking the optimal $k$ where $\rho$ reaches minima.
    Blue and orange curves represent the low-mass and high-mass galaxies, respectively, with corresponding minima indicated by open circles.
    }
    \label{fig:spearman}
\end{figure}

There are two commonly used continuous methods.
The projected radius ($R$) is the most convenient tracer, as it can be easily measured without spectroscopic observations.
Although $R$ is not typically used as a direct tracer of infall time due to the severe projection effect, it remains an effective and widely adopted tool for tracing environmental variations \cite[e.g.,][]{Mahajan_2011,Taranu_2014,Haines_2015,Maier_2019,Lopes_2023}.
Another one is the caustic line. Theoretically, the infall pattern in the phase space associated with a spherically symmetric density perturbation can be described by the caustic profile, which is nearly hyperbolic \citep{Regos_1989,Diaferio_1997}.
Thus, many works have used the hyperbolic caustic profiles in $R-V$ diagrams to quantify the infall process of galaxies \cite[e.g.,][]{Haines_2012,Noble_2013,Noble_2016,Kim_2023}.
Here, we introduce a continuous tracer, the caustic distance, which is defined as:
\begin{equation}
    d_{\rm caustic} = \sqrt {R \times V}
\end{equation}
The reference curves of the projected radii and caustic profiles are plotted in the left and middle panels of Fig.\ref{fig:continuous_map}, respectively.

In addition to these two methods, another natural choice is the linear partition, which is similar to $R$ but uses oblique lines, as illustrated in the right panels of Fig.\ref{fig:continuous_map}.
To quantify linear partitions such as $R$ and $d_{\rm caustic}$, we introduce a new continuous tracer, the linear distance ($d_{\rm linear}$), which is defined as the perpendicular distance from the origin to the oblique lines:
\begin{equation}\label{equ:d_linear}
    d_{\rm linear} = \frac{|V|-k R}{\sqrt{1+k^2}}
\end{equation}
where $k$ is the slope of these lines.
To determine the optimal slope, we test a range of slopes and use the Spearman rank correlation coefficient $\rho$ \citep{spearman_1904} to examine the monotonic correlation between $t_{\rm fall}$ and $d_{\rm linear}$.
A $\rho$ closer to $\pm1$ indicates a stronger monotonic correlation.
The Spearman coefficients as functions of slopes are illustrated in Fig.\ref{fig:spearman}, with minima indicated by vertical lines.
The optimal slopes are $k_{\rm 3D}=-1.8$ for the phase space ($\rho=-0.6$) and $k_{\rm 2D}=-3.7$ for the $R-V$ diagram ($\rho=-0.53$).

We also explore the dependence of the optimal slopes on galaxy stellar mass. 
Using the median stellar mass of $10^{9.8}{\rm M_\odot}$ as the threshold, we divide all galaxies into two subsamples: low-mass and high-mass.
The corresponding Spearman correlation coefficients are plotted against slopes as blue and orange curves in Fig.\ref{fig:spearman}. 
Generally, the low-mass subsample exhibits relatively weaker monotonic correlation and favours a shallower slope.
This trend is similar in both the phase space and the $R-V$ diagram. 
However, the differences in optimal slopes between the two subsamples are small and have very little effect on the resulting correlation coefficients.
Therefore, we adopt the slopes derived from the full galaxy sample, $k_{\rm 3D}=-1.8$ and $k_{\rm 2D}=-3.7$, for calculating $d_{\rm linear}$ in the rest of this paper.  

We preliminarily compare the three methods by overlaying their reference curves on the median $t_{\rm fall}$ distributions in Fig.\ref{fig:continuous_map}, where the phase space is folded to facilitate the application of these methods.
The vertical lines of 3D radius $r$ and projected radius $R$ fail to trace the variation in $t_{\rm fall}$ due to the projection effect.
The caustic profiles align well with the $t_{\rm fall}$ pattern in the phase space, especially for recent-infall galaxies.
However, this alignment weakens in the $R-V$ diagram, where the pattern is slightly different from that in the phase space.
Moreover, the inner caustic profiles are contaminated by some outer galaxies because they extend infinitely along the axes.
In comparison, the linear partition better captures the variation in $t_{\rm fall}$ than does $R$ while also avoiding the inclusion of outer galaxies in the inner regions, such as caustic profiles.

\begin{figure*}[ht]
    \centering
    \includegraphics[width=0.95\linewidth]{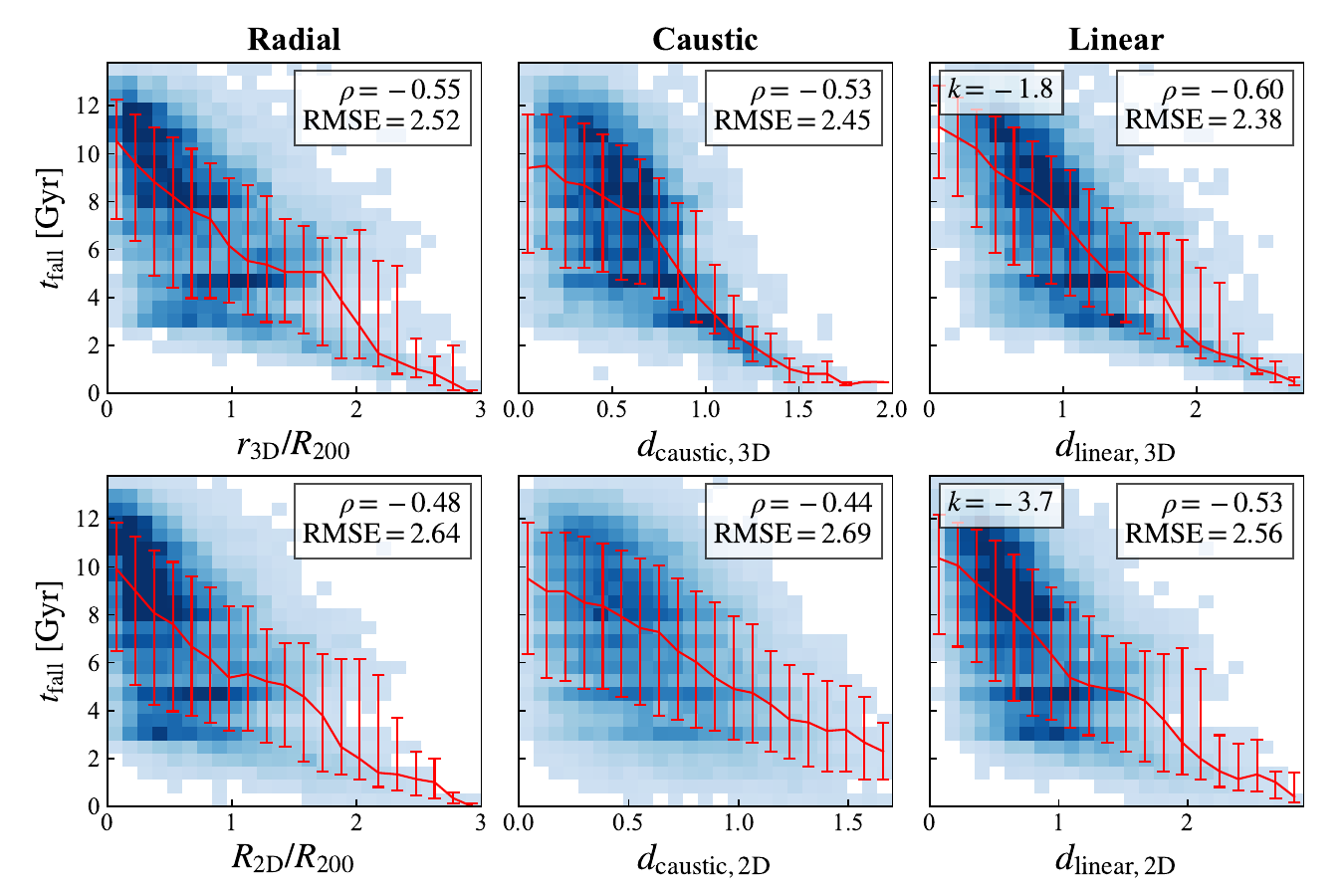}
\caption{Relationships between $t_{\rm fall}$ and the three tracers.
The red curves indicate the median $t_{\rm fall}$ in 20 bins, with error bars indicating the 16th and 84th percentiles.
The Spearman coefficient $\rho$ between $t_{\rm fall}$ and each tracer is written in the corresponding upper right legend, together with the overall $\mathrm{RMSE}$ relative to the bin median.
The slopes for calculating $d_{\rm linear}$ are written in the upper left corners.}
    \label{fig:continuous_relation}
\end{figure*}

We then quantitatively evaluate these methods in Fig.\ref{fig:continuous_relation}, where $t_{\rm fall}$ is plotted against the three tracers ($R$, $d_{\rm caustic}$, and $d_{\rm linear}$).
The Spearman coefficients between $t_{\rm fall}$ and the tracers are written in the legends.
We divide the tracers into 20 bins, and the median $t_{\rm fall}$ values of these bins are plotted as red curves, with error bars representing the 16th and 84th percentiles.
The overall $\mathrm{RMSE}$s relative to these medians are also provided in the legends.

In the phase space (upper panels of Fig.\ref{fig:continuous_relation}), the radius $r$ exhibits a stronger correlation with $t_{\rm fall}$ ($\rho=-0.55$) but also greater dispersion (2.52 Gyr) than those of $d_{\rm caustic,3D}$ ($\rho=-0.53$, $\mathrm{RMSE}=2.45$ Gyr).
The small error bars of the caustic method in the outer region with $d_{\rm caustic,3D}>1$ highlight its effectiveness in estimating the $t_{\rm fall}$ values of recent-infall galaxies.
In addition, the median $t_{\rm fall}$ trend of $r$ exhibits some nonlinearity between $r=1$ and 2.
This arises because galaxies typically experience multiple orbits before virialization, during which their orbital radii fluctuate repeatedly rather than decrease monotonically.
In comparison, the median $t_{\rm fall}$ trend of $d_{\rm caustic,3D}$ is more linear, because caustic profiles closely follow the typical infall trajectories of galaxies.
The linear partition has the strongest correlation ($\rho=-0.6$) and smallest dispersion (2.38 Gyr).
Nonetheless, its median $t_{\rm fall}$ trend still shows some nonlinearity, which is less pronounced than that of $r$ but not fully eliminated as $d_{\rm caustic,3D}$.

In the $R-V$ diagram, all three methods show weaker correlations with $t_{\rm fall}$ and larger dispersions due to the projection effect.
Among them, the caustic method is impacted the most severely.
Its correlation remains lower ($\rho = -0.44$) than that of $R$ ($\rho = -0.48$), but its dispersion rises significantly to 2.69 Gyr, even exceeding that of $R$ (2.64 Gyr).
The outer region with $d_{\rm caustic,2D}>1$ is especially affected according to the error bars.
This is expected because the pattern of recent-infall galaxies in the $R-V$ diagram no longer closely follows the caustic profiles as it does in the phase space (see Fig.\ref{fig:continuous_map}).
In contrast, the oscillatory patterns and non-linear median trends of the other two methods remain similar to those in the phase space.
The linear partition, in particular, still achieves the strongest correlation ($\rho = -0.53$) and the smallest dispersion (2.56 Gyr).
Therefore, despite its simple and empirical nature, the linear partition offers a better balance between the projected radii and caustic profiles.

\subsection{Discrete methods}

\begin{figure*}[htb]
    \centering
    \includegraphics[width=0.99\linewidth]{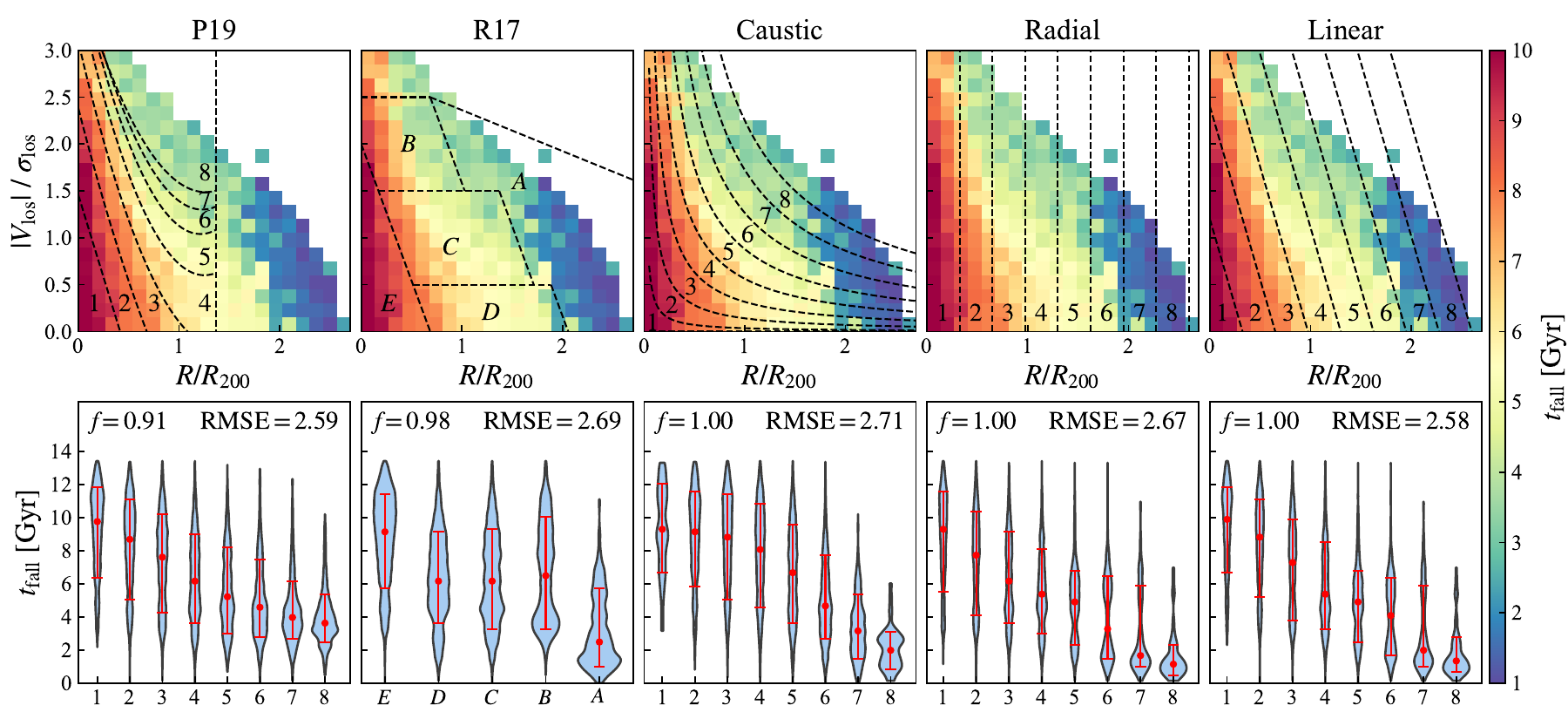}
\caption{Top: Reference curves for each method plotted over the median $t_{\rm fall}$ distribution in the $R-V$ diagram.
Bottom: $t_{\rm fall}$ distributions of individual zones.
The red dots represent the medians, and the error bars represent the 16th and 84th percentiles.
The overall $\mathrm{RMSE}$ values relative to the zone medians are written in the upper right corners.
The fraction of galaxies covered by each method is written in the upper left corner of the corresponding panel.}
    \label{fig:Discrete}
\end{figure*}

Beyond continuous tracers, two studies have attempted to manually divide the $R-V$ diagram into discrete zones, each associated with characteristic infall times.
\citet[][hereafter \citetalias{rhee_2017}]{rhee_2017} used zoom-in hydrodynamic simulations to explore the typical path of galaxies in the $R-V$ diagram, as they infall into the cluster potential.
They divided the $R-V$ diagram into five distinct zones by maximizing the fraction of galaxies belonging to a specific infall population in each zone.
This method has been validated by several studies as effective in tracing galactic evolution in cluster environments \citep[e.g.,][]{Brambila_2023,Ding_2023}.
\citet[][hereafter \citetalias{pasquali_2019}]{pasquali_2019} proposed an eight-zone discretization method based on the mean infall time distribution in the $R-V$ diagram.
Their quadratic curves outperformed both the caustic profiles and the projected radius and have also been widely adopted \citep[e.g.,][]{Sampaio_2021,Oxland_2024,Sampaio_2024}.

Here, we compare the two discrete methods together with three continuous methods in Fig.\ref{fig:Discrete}.
To facilitate comparison, we discretize the continuous tracers into eight equal-width bins between the ranges of $0<R<2.6$, $0<d_{\rm caustic}<1.5$, and $0<d_{\rm linear}<2.6$, as illustrated by the reference curves in the upper panels.
The \citetalias{pasquali_2019} and \citetalias{rhee_2017} discretizations are adjusted from their original formulations, as their projected radii are normalized by $R_{\rm vir}$, whereas we use $R_{200}$ \citep[$R_{200}=0.73R_{\rm vir}$, assuming an NFW profile with concentration $c=4$, ][]{Reiprich_2013}.
The $t_{\rm fall}$ distributions of these zones are shown as violin plots in the lower panels, where red dots represent the medians and error bars represent the 16th and 84th percentiles.
The overall $\mathrm{RMSE}$s relative to these medians are listed in the upper right corners.

Considering the median $t_{\rm fall}$ trends, all methods exhibit monotonically decreasing trends from the inner to outer zones, except for \citetalias{rhee_2017}, which shows a slightly inverted trend in the $B\sim D$ zones.
This may be because \citetalias{rhee_2017} aims to maximize the proportion of a specific population in each zone rather than consider the overall trend of the median infall time.
Additionally, while most median $t_{\rm fall}$ trends decrease to $<2$ Gyr in the outermost zone, the trend of \citetalias{pasquali_2019} flattens in the outer zones (No. $6\sim8$) at $\approx4$ Gyr.
This is because \citetalias{pasquali_2019} excludes the region outside $R_{\rm vir}$ to avoid interlopers.
The fractions of galaxies covered by each method are shown in the upper left corners of the bottom panels.
While other methods cover nearly all member galaxies, \citetalias{pasquali_2019} excludes $\sim10\%$ of them, which are mainly recent-infall galaxies with $t_{\rm fall}<2$ Gyr.
Therefore, this exclusion results in a flattening median $t_{\rm fall}$ trend.

For the dispersions, both \citetalias{pasquali_2019} and the linear partition exhibit the smallest dispersion of $\approx2.6$ Gyr, which is approximately $0.1$ Gyr smaller than those of the other three methods.
This difference is fairly small compared with the dispersions themselves.
As will be discussed in Section \ref{sec:orbit_overlap}, the accuracy of estimating the infall time in the $R-V$ diagram is limited by the intrinsic orbital overlap issue, regardless of whether complex discretization, theory-based caustic lines, or a simple linear partition is used.

Additionally, the fractions of galaxies covered by each method are shown in the upper left corners of the bottom panels.
While other methods cover nearly all member galaxies, \citetalias{pasquali_2019} ignores $\sim10\%$ of them because it covers only the region within $R_{\rm vir}$ to avoid interlopers.
The ignored galaxies are mainly the recent-infall galaxies with $t_{\rm fall}<2$ Gyr.
Therefore, the median $t_{\rm fall}$ trend of \citetalias{pasquali_2019} flattens in the outer zones (No. $6\sim8$) at $\approx4$ Gyr, while the median trends of other methods decrease to $<2$ Gyr in the outer most zone.

\section{Discussion} \label{sec:discussion}

Since estimating the infall times of galaxies with their positions in the $R-V$ diagram is limited by large intrinsic dispersion, here, we explore two potential approaches to improve accuracy and discuss the primary factors contributing to the dispersion.

\subsection{The dynamic states of clusters}\label{sec:clu_dyn}

\begin{figure}[ht]
    \centering
    \includegraphics[width=0.9\linewidth]{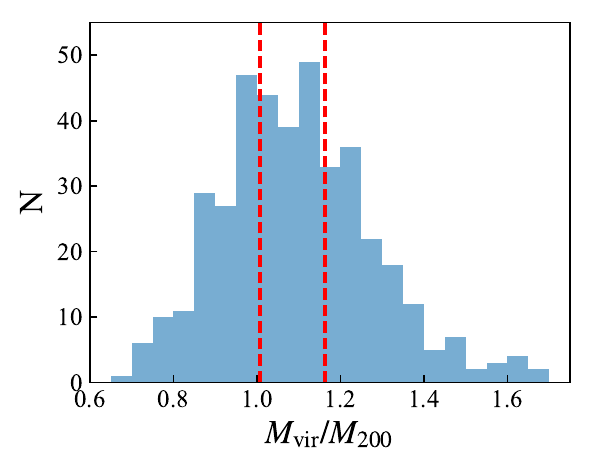}
\caption{ $M_{\rm vir}/M_{200}$ distribution of 408 clusters.
The vertical dashed lines represent the criteria for dividing clusters into three equally sized subsamples.}
    \label{fig:clu_dyn}
\end{figure}
\begin{figure*}[ht]
    \centering
    \includegraphics[width=0.8\linewidth]{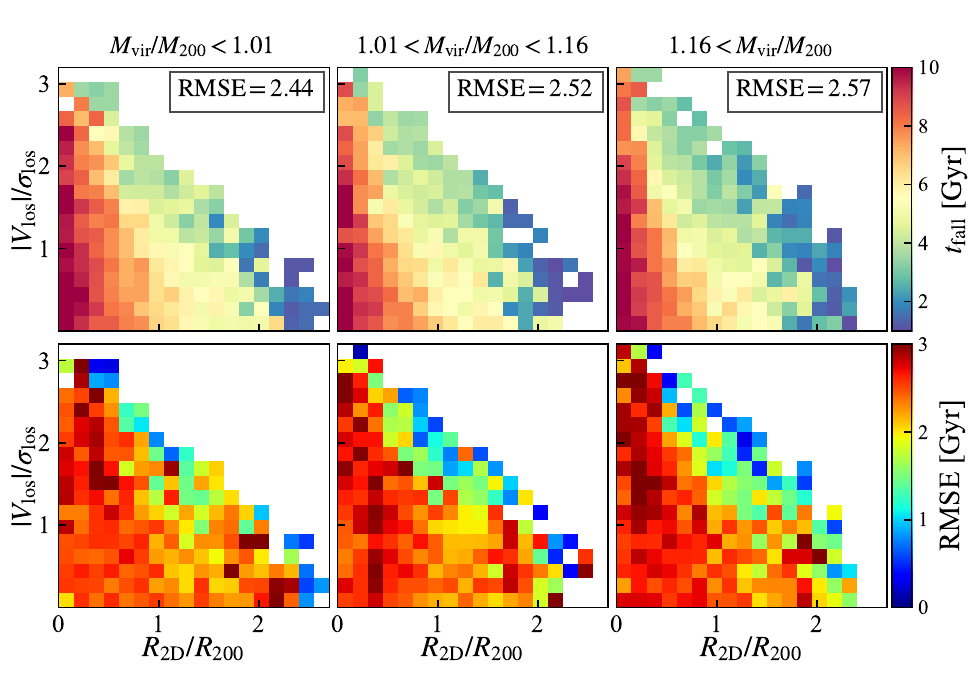}
\caption{Median $t_{\rm fall}$ (top) and $\mathrm{RMSE_{pix}}$ (bottom) distributions of 3 cluster samples in the $R-V$ diagram.
The $M_{\rm vir}/M_{200}$ range of each sample is labelled in the titles.
The overall $\mathrm{RMSE}$s are written in the legends of the top panels.}
    \label{fig:dyn_5sam}
\end{figure*}\begin{figure*}[ht]
    \centering
    \includegraphics[width=0.99\linewidth]{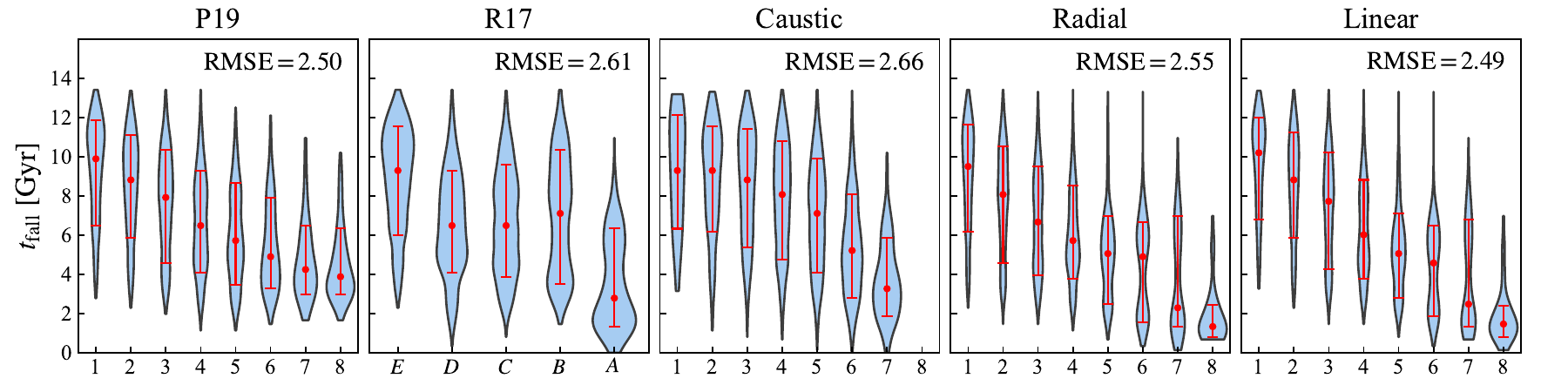}
\caption{Five methods applied to the most relaxed cluster sample, which have $M_{\rm vir}/M_{200}<1.01$.
The elements are the same as those in Fig.\ref{fig:Discrete}.
Note that no galaxy is located in the 8th zone of the caustic profiles; thus, it is empty.}
    \label{fig:relax}
\end{figure*}

The spatial distribution and motion of member galaxies are closely associated with the dynamic state of the host clusters \citep[e.g.,][]{Dressler_1988,Wen_2013}.
Non-relaxed or merging clusters typically exhibit a greater number of infalling galaxies, wherein the motions are significantly disturbed.
In contrast, relaxed and virialized clusters have more regular structures.
Their member galaxies are concentrated towards the centres and have stable motions.
Therefore, here, we explore how the dynamic state of clusters affects the estimation of  infall time in the $R-V$ diagram.

We employ the ratio of the virial mass $M_{\rm vir}$ to $M_{200}$ as an indicator of a cluster's dynamic state.
The virial mass is calculated according to the virial theorem:
\begin{equation}
    M_{\rm vir} = \frac{3R_{200}\ \sigma^2_{\rm los} }{ G}
\end{equation}
The parameters $M_{200}$, $R_{200}$, and $\sigma_{\rm los}$ are all available in the observations.
A virialized cluster has a theoretical mass comparable to its real mass, that is, $M_{\rm vir}/M_{200}\lesssim1$, whereas a less relaxed cluster has a larger $M_{\rm vir}/M_{200}$.
The $M_{\rm vir}/M_{200}$ distribution of our clusters is shown in Fig.\ref{fig:clu_dyn}.
We divide the cluster sample into three equally sized subsets, with the thresholds indicated by the vertical dashed lines.

The median $t_{\rm fall}$ and $\mathrm{RMSE_{pix}}$ distributions in the $R-V$ diagram of the three cluster samples are shown in Fig.\ref{fig:dyn_5sam}, with the overall $\mathrm{RMSE}$s indicated in the legends.
As expected, the dispersion clearly tends to increase with increasing $M_{\rm vir}/M_{200}$.
In the most relaxed sample with $M_{\rm vir}/M_{200}<1.01$, the intrinsic dispersion is $2.44$ Gyr, which is $0.13$ Gyr smaller than that of the least relaxed sample ($M_{\rm vir}/M_{200}>1.16$) and $0.09$ Gyr smaller than that of all clusters together (Fig.\ref{fig:tinf}).
This trend is also evident in the detailed $\mathrm{RMSE_{pix}}$ distributions.
The area with dispersion exceeding 3 Gyr increases in sequence.

We then apply the five methods for estimating infall times to the most relaxed sample, as shown in Fig.\ref{fig:relax}.
The features of these methods remain consistent with their application to the entire galaxy sample (Fig.\ref{fig:Discrete}), but the accuracies are improved by $\approx0.1$ Gyr, in agreement with the 0.09 Gyr improvement in overall $\mathrm{RMSE}$.
Therefore, in observational studies, focusing on relaxed and virialized clusters may help reduce the dispersion in estimating infall times.
Other dynamical indicators \citep[e.g.,][]{Valles_2025} may better probe the dynamical states of clusters.
However, considering the $\mathrm{RMSE}$ difference of only 0.13 Gyr between our least and most relaxed samples shown in Fig.\ref{fig:dyn_5sam}, more precise classification of clusters might not significantly improve the accuracy of infall time estimation.

\subsection{Two estimates}\label{sec:two_esti}

\begin{figure}[ht]
    \centering
    \includegraphics[width=0.9\linewidth]{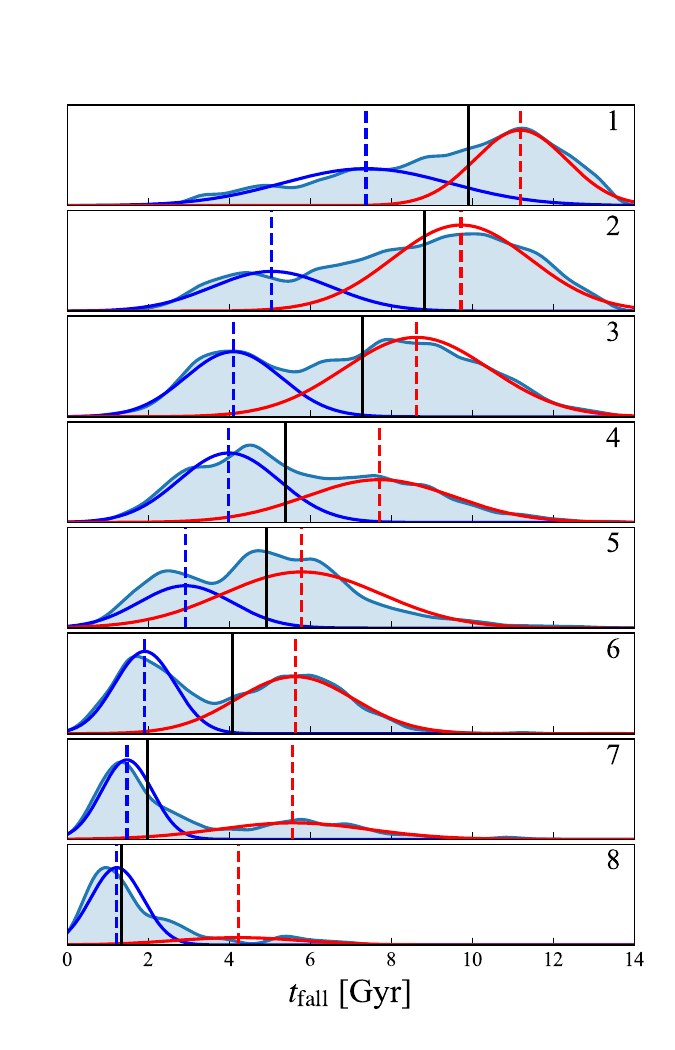}
\caption{Infall time distributions in the 8 zones of the linear partition method.
The zone numbers are indicated in the upper right corners.
The blue shadings represent the $t_{\rm fall}$ distributions constructed via Gaussian kernel density estimation with a kernel size of $0.3$ Gyr.
The black vertical lines indicate the median $t_{\rm fall}$ in each zone.
The blue and red curves are the two components obtained through the GMM, with the blue and red dashed lines indicating their peak locations. }
    \label{fig:two_esti}
\end{figure}

When inspecting the $t_{\rm fall}$ distributions in Fig.\ref{fig:Discrete}, we observe a universal multimodality across them.
A similar phenomenon has also been discussed in previous studies \citep[e.g.,][]{Oman_2013}.
Therefore, employing the median value as a single estimate for infall time would be inherently inaccurate, whether in pixels or in zones.
Taking the linear partition as an example, the $t_{\rm fall}$ distributions in 8 zones are plotted in Fig.\ref{fig:two_esti}, where the black dashed lines indicate the median infall times.
It is clear that in most zones, the median value deviates from any local density peak.
The specific zone ranges and median values are provided in Table \ref{tab:esti}.

Given this multimodality, it is reasonable to consider using multiple estimates rather than a single median value.
For simplicity, we employ two estimates per zone.
We employ the Gaussian mixture model (GMM) to model the $t_{\rm fall}$ distribution in each zone, with the component number fixed as two, implemented with the \texttt{Scikit-learn} \citep{scikit_learn} package in Python.
The fitted Gaussian components are presented as the red and blue curves in Fig.\ref{fig:two_esti}, with peaks indicated by dashed lines.
In most zones, the two peaks provide a better match to the local density peaks in the $t_{\rm fall}$ distributions than the median value does.
The two peaks serve as the two estimates of $t_{\rm fall}$ in each zone, which are listed as $t_1$ and $t_2$ in Table \ref{tab:esti}.

Note that two Gaussian components do not perfectly capture the multimodal nature of the $t_{\rm fall}$ distributions.
For example, at least three peaks are visible in Zone 5.
However, it is more convenient and practical in observations to employ two rather than more estimates.
By considering various physical properties of a galaxy, such as colour, the star formation rate, and the gas fraction, it is possible to roughly determine whether it should have a younger or older infall time.
However, dividing galaxies into more subsamples would be tricky and unreliable.

Assuming that we can reliably divide galaxies into earlier-infall and later-infall populations, we calculate the dispersions when using two estimates.
For each galaxy, we compare its two errors relative to both estimates and use the smaller one to calculate $\mathrm{RMSE}$, which is listed as $\mathrm{RMSE_{two}}$ in Table \ref{tab:esti}.
Compared with the RMSE when the median value is used as the single estimate ($\mathrm{RMSE_{median}}$), the use of two estimates significantly improves the accuracy to $\lesssim1.5$ Gyr, even reaching $\sim1$ Gyr in the outer $6\sim8$th zones.
This improvement clearly demonstrates the advantage of using two infall time estimates in each zone over using one.

\begin{table*}[h]
\centering
\caption{Infall time estimates in 8 zones of the linear partition. } \label{tab:esti}
\begin{tabularx}{0.95\textwidth}{lcccccccc}
\toprule
 Zone & 
 Range \tablefootmark{a} &
 Median $t_{\rm fall}$ \tablefootmark{b} &
 $\mathrm{RMSE_{median}}$ \tablefootmark{c} &
 $t_1$ \tablefootmark{d} & $t_2$ \tablefootmark{d} &
 $\mathrm{RMSE_{two}}$ \tablefootmark{e} &
 $f_1:f_2$ \tablefootmark{f} &
 $f_{\rm intl}$  \tablefootmark{g}  \\
\midrule
1 & $ 0.000<d_{\rm linear}<0.325$ & 9.90 & 2.57 & 7.37 & 11.18 & 1.40 & 0.47:0.53 & 0.00  \\
2 & $ 0.325<d_{\rm linear}<0.650$ & 8.82 & 2.71 & 5.04 & 9.72 & 1.47 & 0.29:0.73 & 0.01  \\
3 & $ 0.650<d_{\rm linear}<0.975$ & 7.28 & 2.70 & 4.09 & 8.62 & 1.44 & 0.35:0.65 & 0.03  \\
4 & $ 0.975<d_{\rm linear}<1.300$ & 5.37 & 2.45 & 3.99 & 7.70 & 1.32 & 0.52:0.48 & 0.07  \\
5 & $ 1.300<d_{\rm linear}<1.625$ & 4.90 & 2.20 & 2.92 & 5.78 & 1.43 & 0.31:0.69 & 0.18  \\
6 & $ 1.625<d_{\rm linear}<1.950$ & 4.08 & 2.19 & 1.91 & 5.62 & 1.08 & 0.42:0.58 & 0.37  \\
7 & $ 1.950<d_{\rm linear}<2.275$ & 1.98 & 2.58 & 1.47 & 5.56 & 1.09 & 0.62:0.38 & 0.60  \\
8 & $ 2.275<d_{\rm linear}<2.600$ & 1.34 & 1.54 & 1.22 & 4.21 & 0.81 & 0.81:0.19 & 0.85  \\
\bottomrule
\end{tabularx}
\tablefoot{
\tablefoottext{a}{The range of each zone. $d_{\rm linear}$ is calculated through Equation \ref{equ:d_linear}.}
\tablefoottext{b}{The median infall time in each zone, in Gyr.}
\tablefoottext{c}{The $\mathrm{RMSE}$ relative to the median infall time in each zone, in Gyr.}
\tablefoottext{d}{The two infall time estimates obtained by GMM, in Gyr.}
\tablefoottext{e}{The $\mathrm{RMSE}$ calculated with the smaller one of the two errors relative to the two estimates in each zone, in Gyr.}
\tablefoottext{f}{The relative proportion of two Gaussian components given by GMM.}
\tablefoottext{g}{The fraction of interlopers in each zone.}     }
\end{table*}

\subsection{Orbital overlap}\label{sec:orbit_overlap}

Throughout this paper, we have emphasized the large intrinsic dispersion of $t_{\rm fall}$ in the $R-V$ diagram. However, what causes this dispersion?

During a single-infall orbit, there is a nearly one-to-one correspondence between the $t_{\rm fall}$ of a galaxy and its position in the $R-V$ diagram.
However, galaxies experience multiple orbits before being virialized, which is indicated by the oscillation of $r$ with $t_{\rm fall}$, as shown in Fig.\ref{fig:continuous_relation}.
Consequently, galaxies in different orbits, of course with different $t_{\rm fall}$, overlap with each other in the $R-V$ diagram.
In other words, each position in the $R-V$ diagram corresponds to a complex mixture of galaxies in multiple orbital phases, naturally leading to the multimodality of the $t_{\rm fall}$ distribution discussed in Section \ref{sec:two_esti}.
Here, we further illustrate the orbital overlap issue.

\begin{figure*}[htb]
    \centering
    \includegraphics[width=0.99\linewidth]{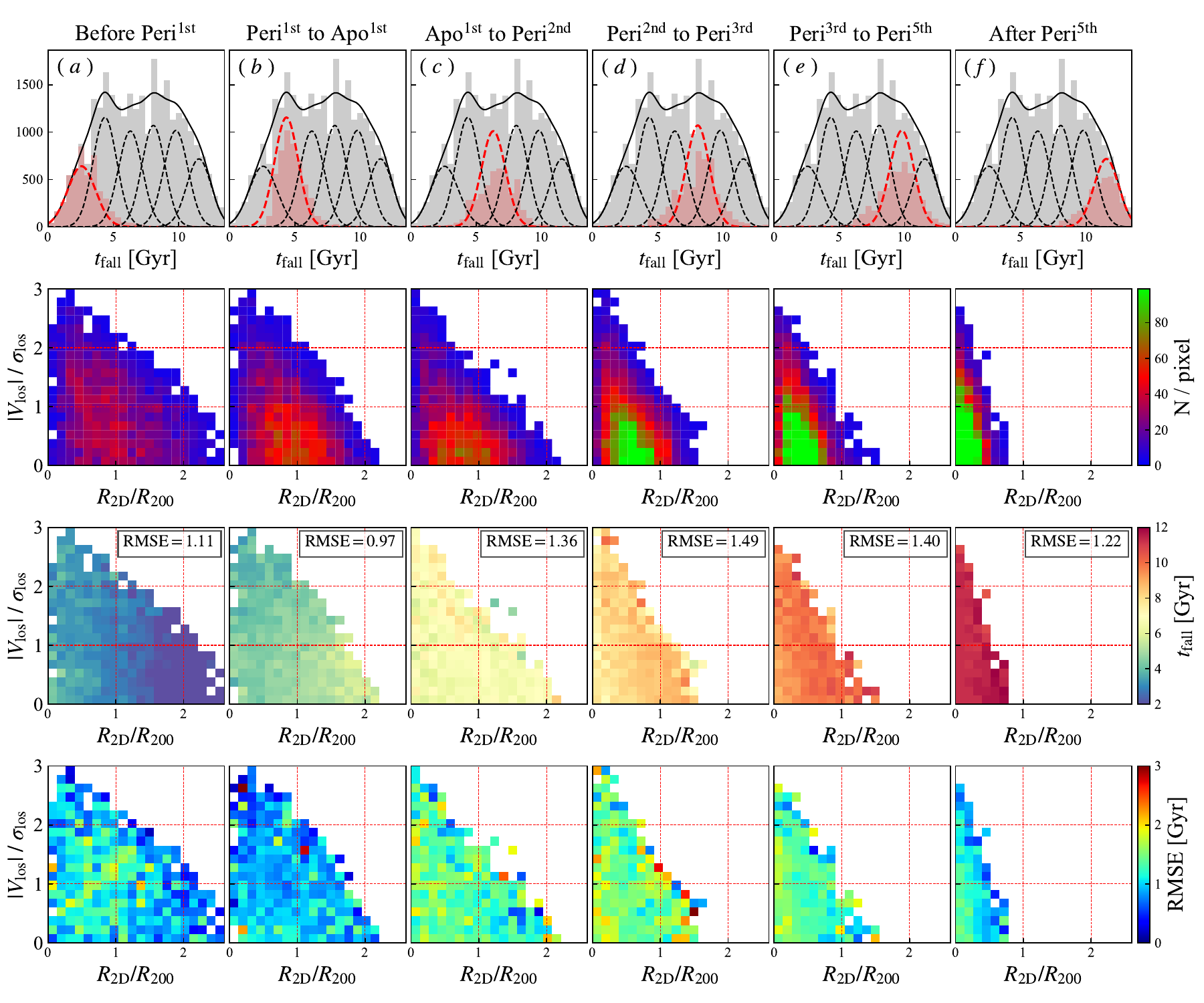}
    \caption{Top panels: Infall time distributions.
    The grey histograms represent all the galaxies, and the dashed and solid curves represent the results of GMM fitting, which yields six Gaussian components.
    The $t_{\rm fall}$ distributions of the six orbital populations are shown as red histograms.
    These curves roughly align with the six Gaussian components, which are highlighted as red dashed curves one by one for visual clarity.
    The 2nd to 4th rows: Number distributions, median $t_{\rm fall}$ distributions, and $\mathrm{RMSE_{pix}}$ distributions in the $R-V$ diagram of the six orbital populations.
    }
    \label{fig:Orbit_overlap}
\end{figure*}

First, the overall $t_{\rm fall}$ distribution for all the galaxies is also multimodal.
We employ the GMM to model the overall $t_{\rm fall}$ distribution, and this time, we free the component number and determine the optimal model through the Bayesian information criterion (BIC).
The best model has six components, and the results are shown in the top panels of Fig.\ref{fig:Orbit_overlap}.
The grey histogram shows the overall $t_{\rm fall}$ distribution, and the six dashed curves represent six Gaussian components.

Moreover, we divide galaxies into 6 populations on the basis of their orbital phases at $z=0$:
\begin{itemize}
\item ({\em a}) Before the 1st pericentre;
\item ({\em b}) From the 1st pericentre to the 1st apocentre;
\item ({\em c}) From the 1st apocentre to the 2nd pericentre;
\item ({\em d}) From the 2nd pericentre to the 3rd pericentre;
\item ({\em e}) From 3rd pericentre to 5th pericentre;
\item ({\em f}) After the 5th pericentre.
\end{itemize}
These populations are arbitrarily determined to ensure that their $t_{\rm fall}$ distributions, shown as the red histograms in the top panels of Fig.\ref{fig:Orbit_overlap}, roughly align with the 6 Gaussian components.
While this alignment is not exact, it still reflects the mixture of different orbital populations in the overall galaxy sample.
The number distributions, median $t_{\rm fall}$ distributions, and $\mathrm{RMSE_{pix}}$ distributions of these 6 populations in the $R-V$ diagram are shown in the second, third, and fourth rows of Fig.\ref{fig:Orbit_overlap}, respectively.

Considering the number distributions, population ({\em a}) galaxies, which are in their first infalling orbits, are spread over the entire $R-V$ diagram.
Some of them even sink into the innermost region rather than remaining in the outer regions.
As the galaxies undergo subsequent orbits, the orbital radii shrink due to dissipation, and galaxies with longer orbital histories occupy the region with a smaller $R$ value (see populations \textit{d}, \textit{e}, \textit{f}).
Therefore, the inner region of the $R-V$ diagram mixes more orbital populations than the outer region does, leading to greater dispersion in the inner region, as shown in Fig.\ref{fig:tinf}.
This also explains the reduced dispersion in more relaxed clusters (Section \ref{sec:clu_dyn}): relaxed clusters contain fewer population ({\em a}) galaxies, which occupy a larger area of the $R-V$ diagram.

For the median $t_{\rm fall}$ distributions in the third row of Fig.\ref{fig:Orbit_overlap}, the trend of population ({\em a}) remains consistent with that of all galaxies in Fig.\ref{fig:tinf}: a larger $t_{\rm fall}$ in the inner region.
However, the population ({\em b}) exhibits the opposite trend, with a larger $t_{\rm fall}$ at a larger $R$.
These galaxies are moving outwards and are usually referred to as `splashback galaxies'.
The opposite trends from these outwards-bound orbital phases further complicate the relationship between $t_{\rm fall}$ and the positions in the $R-V$ diagram.
In the subsequent populations, the median $t_{\rm fall}$ varies little with position, probably because these galaxies are close to virialization and exhibit more isotropic motion.

Focusing on the dispersions, the overall $\mathrm{RMSE}$s of these populations are reported in the legends of the third-row panels of Fig.\ref{fig:Orbit_overlap}.
Each population has a dispersion of $<1.5$ Gyr, and particularly, populations ({\em a}) and ({\em b}) have dispersions of $\approx1$ Gyr.
Therefore, the overall dispersion of 2.53 Gyr results from the superposition of smaller dispersions of these orbital populations.
The detailed $\mathrm{RMSE_{pix}}$ distributions in the bottom panels of Fig.\ref{fig:Orbit_overlap} are fairly uniform across the $R-V$ diagram, and their values are consistent with the overall $\mathrm{RMSE}$s of the corresponding populations.
Therefore, the complex patterns in the $\mathrm{RMSE_{pix}}$ distributions and relative dispersion distributions of all the galaxies (Fig.\ref{fig:tinf}) are also established by the superposition of these orbital populations.

These results highlight the dominant contribution of orbital overlap to the dispersion of the infall time estimation.
The degeneracy in orbital phases remains challenging to resolve using observational data alone.
As discussed in Section \ref{sec:two_esti}, incorporating the additional physical properties of galaxies may help partially mitigate this degeneracy, although further studies are needed to confirm its effectiveness.

\subsection{Other sources of dispersion} \label{sec:secondary_dis}

In addition to orbital overlap, several secondary sources contribute to dispersion.

The projection effect plays a significant role.
As shown in Fig.\ref{fig:infall_his}, the infall trajectory is clear in the phase space but is obscured in the $R-V$ diagram by the projection effect.
All three continuous methods perform better when applied in the phase space than in the $R-V$ diagram, as illustrated in Fig.\ref{fig:continuous_relation}.
In observational studies, measuring the distance and two projected velocity components is often extremely challenging, making the projection effect largely unsolvable.

Many galaxies infall into clusters as groups, which have their own internal dynamics \citep{Diaferio_2009,Serra_2011}.
These motions can introduce intrinsic and random noise in the $R-V$ diagram.
However, with spectroscopic data and a sophisticated clustering algorithm, it is still possible to identify substructures within clusters and treat them separately \citep{Yu_2015, Qu_2023}.

\subsection{Interlopers}

\begin{figure}[htb]
    \centering
    \includegraphics[width=0.99\linewidth]{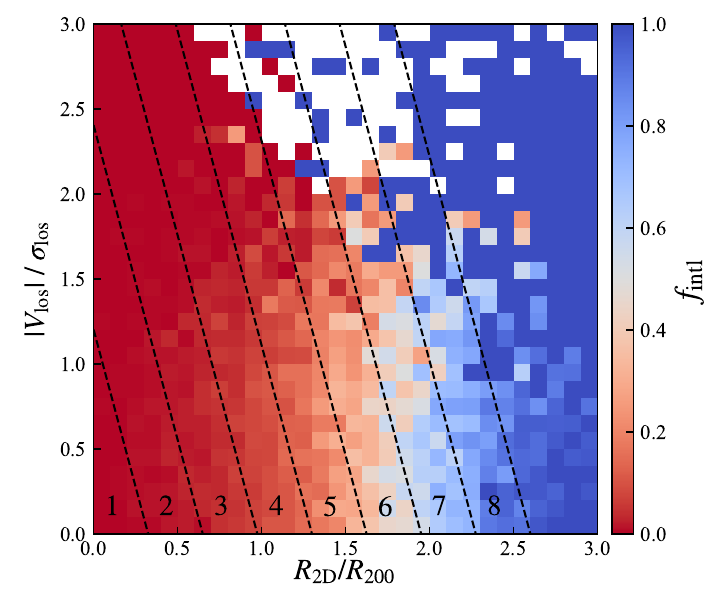}
    \caption{Fraction of interlopers in the $R-V$ diagram.
    Reference lines of the linear partition method are plotted as dashed lines.}
    \label{fig:Interloper}
\end{figure}

In observation, the projection effect not only causes dispersions in infall time estimation, but also introduces interlopers, which are not physically associated with the cluster but lie within the $R-V$ diagram.
Since our analysis in this paper is limited to true member galaxies, we here assess the potential impact of interlopers.

We define interlopers as galaxies that are not classified as members by TNG group catalogues but are located within $10R_{200}$ around our clusters.
Galaxies beyond this distance have little chance of appearing where the cluster members are located in the $R-V$ diagram.
Fig.\ref{fig:Interloper} shows the fraction of interlopers in the $R$–$V$ diagram, along with the reference lines of the linear partition method.
It is clear that interlopers mainly dominate the outer region with $R\gtrsim2$, roughly around zone 7, zone 8 and farther outward.
Similar results have been reported in previous works \citep[e.g.,][]{Oman_2016}.
The interloper fractions in eight linear zones are listed as $f_{\rm intl}$ in Table \ref{tab:esti}.
As expected from Fig.\ref{fig:Interloper}, interlopers constitute more than 50\% of the galaxies in zones 7 and 8, whereas inner regions are dominated by true members.
We therefore caution that our infall time estimates in zone 7 and zone 8 may be unreliable in observation and should be interpreted with care.

\section{Summary} \label{sec:summary}

Accurately estimating the infall time of galaxies infalling into clusters is crucial for understanding the environmental evolution of galaxies, yet achieving this estimation is challenging in observations.
In this work, we utilize the TNG300-1 simulation to systematically evaluate five methods for estimating the infall time in the $R-V$ diagram, including the projected radii, caustic profiles, two discrete zoning methods, and a simple linear partition method.

These five methods exhibit distinct characteristics.
Caustic profiles perform well for recent-infall galaxies in the phase space but are less effective in the $R-V$ diagram because of the projection effect.
The projected radii $R$, which are also affected by the projection effect, perform better than the caustic profiles do in the $R-V$ diagram.
Of the two discretizations, \citetalias{pasquali_2019} has the smallest dispersion ($\approx2.6$ Gyr), although it excludes 10\% of the infalling galaxies, whereas \citetalias{rhee_2017} shows slight non-monotonicity in the median $t_{\rm fall}$ trend.
Compared with the other methods, the simple linear partition performs slightly better, showing the same smallest dispersion as \citetalias{pasquali_2019} and a stronger correlation with infall time compared with the projected radii and caustic profiles.

However, the difference between the accuracies of the various methods is only $\approx0.1$ Gyr, which is negligible compared with the overall dispersion ($\gtrsim2.6$ Gyr).
In other words, these methods are all limited by the intrinsic dispersion of the infall time (2.53 Gyr), irrespective of their specific strategies for partitioning the $R-V$ diagram.
Given this limit, we explore two potential approaches for improving accuracy.
First, more relaxed clusters exhibit less dispersion overall, allowing for more accurate estimation.
Second, considering the multimodality of the infall time distributions, using two estimates of the infall time in each zone, rather than a single median, improves the accuracy to $\lesssim1.5$ Gyr.
In observations, the appropriate choice between smaller or larger infall times could be guided by various physical properties of the galaxy.
Both the single-estimate and two-estimate values and corresponding dispersions in 8 zones of the linear partition are listed in Table 1, which may serve as a quantitative reference for future studies.

Finally, we demonstrate the orbital overlap issue that primarily contributes to the large intrinsic dispersion.
The multimodal infall time distributions of all the galaxies are decomposed into six Gaussian components via the GMM.
These components roughly align with six populations of galaxies in different orbital phases.
The six orbital populations overlap with each other in the $R-V$ diagram, leading to orbital degeneracy among galaxies and consequently large dispersion in infall times.
Additionally, secondary factors such as the projection effect and internal dynamics of infalling groups contribute to the overall dispersion.

Our results underscore the challenges and limitations of estimating infall times from observational data.
Nevertheless, linear partitioning provides a simple and robust empirical framework for statistically linking the evolution of galaxies with their infall process.
With upcoming surveys, such as DESI \citep{Levi_2013,DESI_2016a} and Euclid \citep{Laureijs_2011,Euclid_2024}, these methods will provide valuable insights into how the cluster environment influences galaxy evolution during the infall process.

\begin{acknowledgements}
We thank Lizhi Xie for providing constructive comments and valuable advice.
We thank the IllustrisTNG collaboration for providing free and convenient access to the data used in this work, which can be accessed at \url{www.tng-project.org}.
This work was made possible via several open-source software packages: AstroPy \citep{astropy}, Matplotib \citep{matplotlib}, NumPy \citep{numpy}, Pandas \citep{pandas}, SciPy \citep{scipy} and \texttt{Scikit-learn} \citep{scikit_learn}.
\end{acknowledgements}

% \bibliography{references}{}
% \bibliographystyle{aa}

\end{document}